# Third Harmonic Enhancement Harnessing Photoexcitation Unveils New Nonlinearities in Zinc Oxide

Soham Saha[1*], Sudip Gurung[2], Benjamin T. Diroll[1], Suman Chakraborty[1], Ohad Segal[3], Mordechai Segev[3], Vladimir M. Shalaev[4, 5], Alexander V. Kildishev[4,5], Alexandra Boltasseva[4,5] and Richard D. Schaller[1,6]

[1] *Argonne National Laboratory, Lemont, IL 60439, USA*

[2] *CREOL, the College of Optics and Photonics, Orlando, FL 32816, USA*

[3] *Solid State Institute, Technion–Israel Institute of Technology, Haifa 32000, Israel*

[4] *School of Electrical and Computer Engineering, Birck Nanotechnology Center, Purdue University, West Lafayette, IN, USA*

[5] *Purdue Quantum Science and Engineering Institute, Purdue University, West Lafayette, IN, USA*

[6] *Department of Chemistry, Northwestern University, Evanson, IL, USA*

*\*sohamsaha@anl.gov*

## Abstract

Nonlinear optical phenomena are at the heart of various technological domains such as high-speed data transfer, optical logic applications, and emerging fields such as non-reciprocal optics and photonic time crystal design. However, conventional nonlinear materials exhibit inherent limitations in the post-fabrication tailoring of their nonlinear optical properties. Achieving real-time control over optical nonlinearities remains a challenge. In this work, we demonstrate a method to switch third harmonic generation (THG), a commonly occurring nonlinear optical response. Third harmonic generation enhancements up to 50 times are demonstrated in zinc oxide films via the photoexcited state generation and tunable electric field enhancement. More interestingly, the enhanced third harmonic generation follows a quadratic scaling with incident power, as opposed to the conventional cubic scaling, which demonstrates a previously unreported mechanism of third harmonic generation. The THG can also be suppressed by modulating the optical losses in the film. This work shows that the photoexcitation of states can not only enhance nonlinearities, but can create new processes for third harmonic generation. Importantly, the proposed method enables real-time manipulation of the nonlinear response of a medium. The process is switchable and reversible, with the modulations occurring at picosecond timescale. Our study paves the way to boost or suppress the nonlinearities of solid-state media, enabling robust, switchable sources for nonlinear optical applications.

## Introduction

Nonlinear optics, the field of light-matter interactions under high-light-intensity conditions, has emerged as a critical field with a plethora of applications in signal processing, telecommunications, biomedical imaging, and quantum information science[1]. Among various nonlinear optical phenomena, third harmonic generation (THG) stands out as a very basic process where three photons of frequency $\omega$ combine to generate a single photon with tripled frequency[2] ($3\omega$). THG has numerous applications in material characterization[3], imaging[4], microscopy[5], and the development of advanced optical devices.

Transparent conducting oxides (TCOs)[6,7], with their unique combination of low optical losses and high electrical conductivity, play a pivotal role in nonlinear optical applications ranging from tunable optics[7] to photonic time crystal design[8,9]. Their unique epsilon-near-zero properties also allow for effects ranging from giant optical nonlinearity enhancements to on-chip modulator design[10–19]. Zinc oxide (ZnO) has garnered considerable attention as a promising TCO[20]. It exhibits excellent optical properties with low optical losses[20], large laser damage threshold[21], optical tailorability when doped with various dopants[22,23] making it an ideal candidate for optical devices[20,24,25]. Active modulation of the nonlinear optical properties of various TCOs has already resulted in on-chip modulators[26], metasurfaces[27], and ultrafast switches[28,29], opening new directions in optical data transfer devices[30,31]. Thus, the prospect of actively tuning the nonlinear response of ZnO opens avenues for developing novel devices with enhanced performance.

Prior methods of controlling the third harmonic generation from various materials have included doping the oxides with metal ions, pushing them to the epsilon-near-zero regime[12], complex nanostructuring schemes to boost the nonlinearities[32,33], and electrical modulation of the free carrier concentration[34]. Our method of boosting THG involves a simpler, fabrication-free scheme, which intriguingly, introduces a new method of third harmonic generation.

This paper presents a unique approach to control third harmonic generation in zinc oxide. The interband optical pumping excites electrons from the valence band to the conduction band, modifies the free-carrier concentration in the film, and consequently, the nonlinear optical response. We actively manipulate the third harmonic generation in zinc oxide films not only via control of free carrier concentration and electrical field-enhancement, but by opening an alternative mode of third harmonic conversion via two-photon absorption. This method enhances THG by an order of magnitude, and, under special conditions, can also suppress the THG. Moreover, the THG enhancement and suppression are both switchable, occurring in the picosecond timescale. This work contributes to the fundamental understanding of nonlinear behavior in the photoexcited state of ZnO and introduces a novel avenue for harnessing and controlling harmonic generation in transparent conducting oxides.

# Results

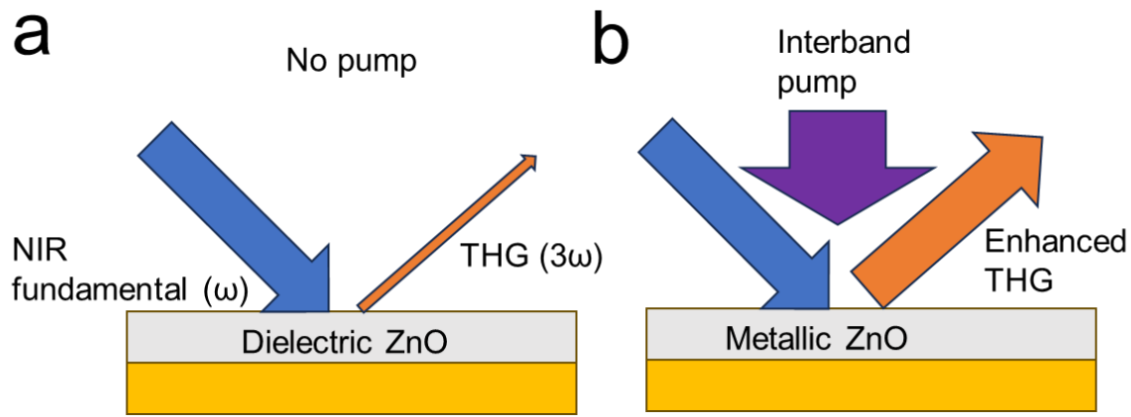

Figure 1. **Basic principle for tunable third harmonic generation.** With a near infrared seed beam, third harmonic generation results in the formation of visible light. Upon pumping the zinc oxide with an interband light-pulse, the generation of the third harmonic beam can be greatly enhanced (or suppressed).

We characterized the third harmonic generation coming from zinc oxide in the specular reflectance direction. Figure 1 shows the basic experimental concept where, without a pump, the collected third harmonic light scales with the cube of the incident near-infrared probe beam, as expected for conventional third harmonic generation. Upon excitation by an optical pump, proximal to the maximum pump-probe overlap, THG is enhanced by up-to 50 times. At greater pump-probe time delays, THG begins to be suppressed as the material relaxes. More interestingly, the pumped third harmonic scales with the square of the intensity of the probe beam at the fundamental frequency $\omega$. The THG scaling with the pump is also more complex: *s*- and *p*-polarized light show significantly different power dependencies, with distinct levels of enhancement and different saturation pump-fluences.

The demonstrated method of THG efficiency control is broadband, does not involve the use of complex nanopatterning, and has the potential of even stronger enhancements employing thinner films utilizing the epsilon-near-zero[18,35,36] and Berreman modes[37,38]. Thus, in addition to opening new routes for THG from robust, laser-tolerant materials employing a simple pump-probe spectroscopy setup, this study also provides a simple way of boosting or suppressing the nonlinearities in optical media.

**Carrier dynamics in zinc oxide films**

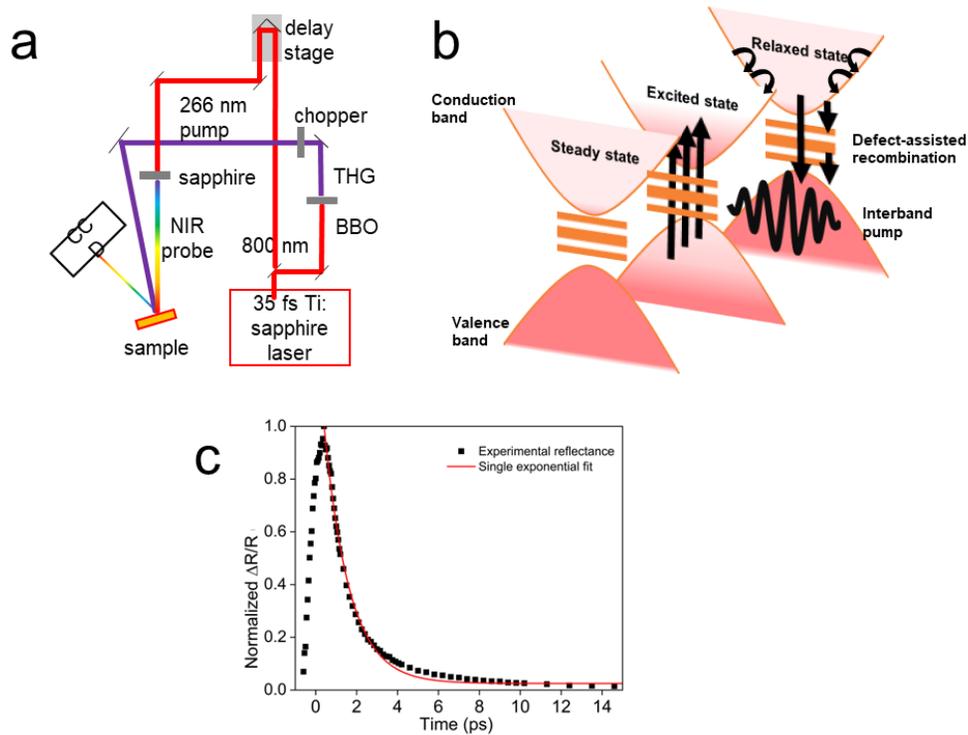

Figure 2. **Optical characterization of carrier dynamics in ZnO films.** a. Experimental setup for transient reflectance spectroscopy b. The photocarrier excitation and recombination process c. Normalized reflectance modulation versus the time (pump-probe delay) for photoexcited ZnO.

In our experiments, first, we studied the steady-state optical properties of a zinc oxide film grown by pulsed laser deposition and its third harmonic generation. A 250 nm film of zinc oxide was grown on an optically thick, epitaxial titanium nitride (TiN) film. TiN was chosen as a backreflector because of its reflectivity in the desired wavelength range and its high laser tolerance[39–42]. The films were characterized by spectroscopic ellipsometry, and their optical properties were fit using a Drude-Lorentz model. Zinc oxide acts as a dielectric, and titanium nitride as a reflective metal in the visible to near-infrared spectral region. Supplementary information S1 has additional details on thin film growth and characterization.

Fig. 2a shows the transient reflectance spectroscopy setup. A 266 nm pump is generated by passing an 800 nm, 35fs, 1kHz repetition rate laser through both doubling and sum frequency BBO crystals. The probe beam at the fundamental frequency $\omega$ is formed by passing the 800 nm through a sapphire crystal, forming a supercontinuum that covers the near-infrared wavelength range. The time delay between the pump and the probe is controlled utilizing a retroreflector on a delay stage, which is used to adjust the path length of the pump.

The carrier dynamics of conducting oxides under an interband pump has been well-studied[24,43–45]. An interband pump generates free carriers, which change the optical properties of the films via Drude dispersion (Fig. 2b), resulting in a change in the reflectance and transmittance. The material relaxes into its unpumped state via defect-assisted Shockley Read Hall recombination[24] after which the optical properties revert to the initial state. The modulation amplitude depends on the pump power that controls the concentration of the photoexcited free-carriers, and the relaxation time depends on their recombination rate. In ZnO films grown under identical conditions, the relaxation time was previously reported to be tens of picoseconds. Under similar excitation, TiN demonstrated a nanosecond response time as shown in prior work[24,46].

The minimum attainable switching time in interband modulation largely depends on the relaxation time of the carriers. The relaxation is best approximated with a single exponential model, given by

$$\left(\frac{\Delta R}{R}\right)(t) = Ae^{-\frac{t}{\tau}} + B \tag{1}$$

where $A$ is the amplitude of the modulation, $t$ is the pump-probe delay time, $\tau$ is the decay constant, and $B$ is an offset attributed to thermal energy dissipation due to lattice heating. The reflectance modulation had a relaxation time of around 10 ps (Fig. 2c), with a fitted 1.26 ps decay constant. The position of the stage for the maximum reflectance modulation is also noted for the third harmonic modulation studies. For the subsequent THG studies, the readings are taken at a 1ps delay after the maximum pump-probe overlap, to prevent the effects from the pump's direct interaction with the probe.

The following sections discuss the enhancement of third harmonic generation efficiency arising from the interband pumping, the dynamic properties of the THG modulation, and the mechanisms that cause the modulation of nonlinearities.

**Third harmonic generation in zinc oxide films**

Preliminary studies of steady-state, third harmonic generation are performed by shining an 1850 nm, 35 fs, laser pulse with a 1 kHz repetition rate onto the zinc oxide surface at 50 degrees, which forms a strong probe beam, with a measured beam diameter of 70 microns. A CCD monitors the THG generated from the surface. A 266 nm interband pump excites photocarriers in the zinc oxide.

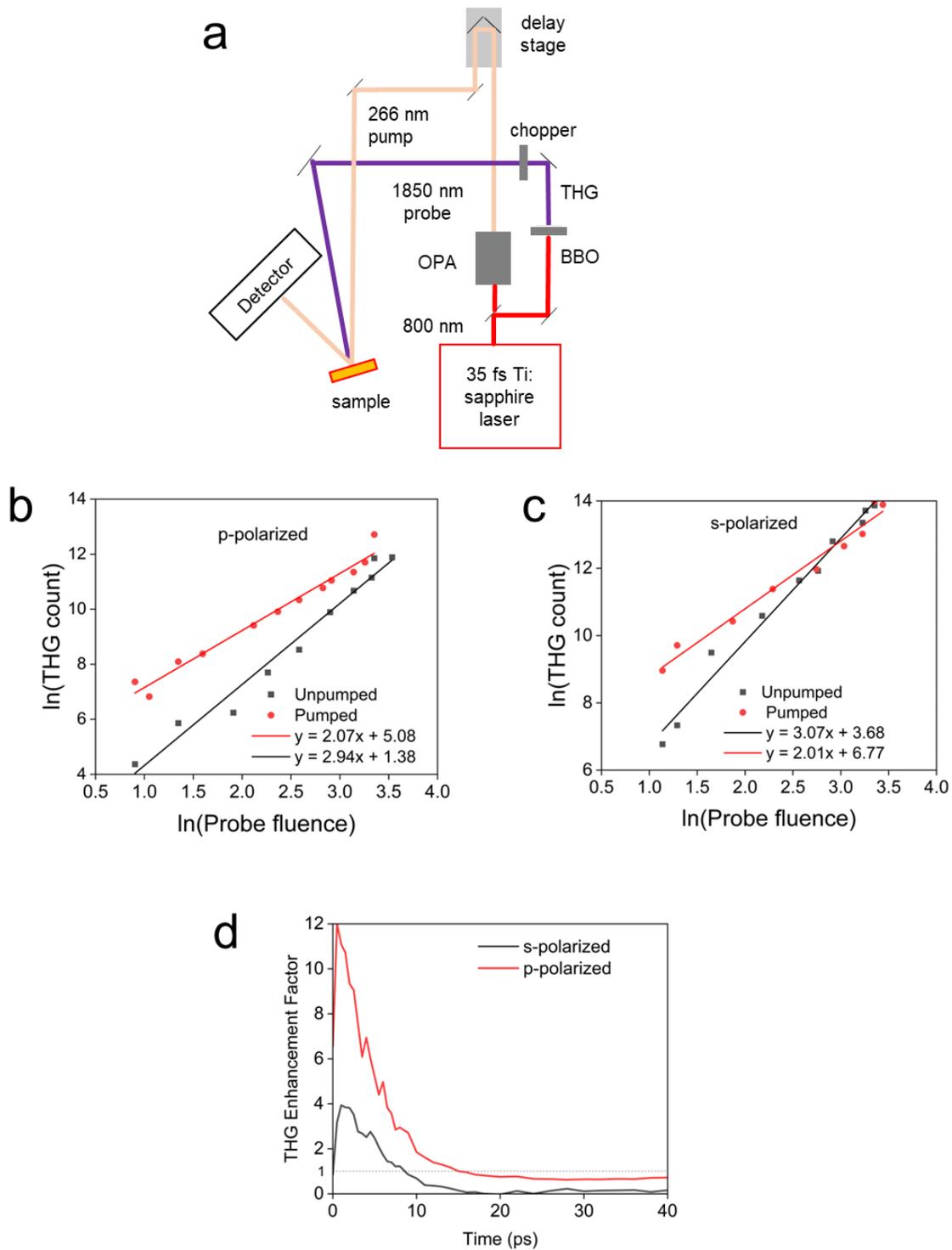

Figure 3 a. Third harmonic measurement setup b. Power dependence (log-scale) of third harmonic generation without pump and c. with pump d. Temporal evolution of third harmonic generation

First, the pump is blocked, the high-fluence probe illuminates the sample surface, and the CCD measures the generated THG light at different probe fluences. In the unpumped state, the extracted third harmonic signal shows a third order power dependence with the near-infrared probe power (black graphs in Fig. 3b,c). Light with s-polarization shows a greater third harmonic efficiency than p-polarized light under the same experimental conditions. This polarization dependence is primarily due to greater field penetration of the parallel component of the electric field into the zinc oxide[47,48]. For s-polarized light, the field is continuous ($E_{\|air} = E_{\|ZnO}$), resulting in greater efficiency of third harmonic generation. As the permittivity of zinc oxide is higher than that of air, the perpendicular component of the electric field is lower inside the film in the unpumped state. Supporting Information Section 3 shows the different field distributions at 3 distinct pump fluences.

We irradiated the films with a pump beam of 3 mW (5.6 mJ/cm$^2$) with a diameter of 260 microns, and investigated the probe power dependence of the films. Interestingly, under influence of the pump, the THG power dependence changes from the third-order to a second-order dependence (Fig. 3b,c, red curves). In these experiments, the probe is delayed by 1 ps after the pump excites the photocarriers, to prevent direct interaction between the pump and the probe pulses. The enhancement, therefore, occurs from the probe's interaction with the photoexcited carriers and lattice in the material. Both s- and p- polarized light show third-harmonic enhancements at this delay, when most of the photoexcited carriers are still in the conduction band. With a probe intensity of 2.5mJ/cm$^2$, s- polarized light shows an enhancement of around 20 times. P-polarized light shows an even larger enhancement, because of the additional field-enhancement resulting from the decreasing permittivity of the oxide.

The enhancement also shows an optimal point with the maximum ratio of output/input, after which the enhancement decreases. For s-polarized light, additionally, a probe power is reached where no additional enhancement is seen, and instead, the THG is observed to be suppressed with additional pump intensity.

Next, we studied the temporal response of the third harmonic enhancement with a fixed pump-probe configuration. A constant pump of 4.3 mJ/cm² was used, with the probe power held at 12.1 mJ/cm². In this configuration, p-polarized THG shows an enhancement of around 12 times with the maximum pump-probe overlap, and s-polarized light shows an enhancement of 4-times (Fig. 3d). With increasing pump-probe delay, the enhancement decreases as the photoexcited carriers recombine, primarily because of the decreasing number of photoexcited states in the conduction band. After a 10-15 temporal delay, we observed a suppression of THG, possibly due to suppression of photon generation because of lattice heating.

## Discussion

THG can be achieved through various optical mechanisms. The most common method involves third-order nonlinear processes within a material, where three incident photons combine to produce a single photon at triple the frequency. This process typically follows a cubic power dependence on the incident intensity due to the involvement of three photons[2].

However, in our pump-modified THG experiments, we clearly observe a 2$^{nd}$ order dependence of the THG efficiency with probe power. A possible explanation is presented in Fig. 4a. For conventional THG in ZnO, electrons from the lowest occupied energy states in the ZnO conduction band absorb three photons, and then recombine, emitting a single photon with 3x the energy. In the pumped-state, the pre-excited electrons are able to absorb two photons of frequency ω, and emit a photon of 3ω when returning to the lowest available states. This results in the square dependence of the THG with respect to probe power[49]. With increasing pump, the number of photoexcited electrons increases, resulting in a direct enhancement of the third harmonic generation (Fig. 4b). However, in this case, the absorptive losses of the film also increase with increasing photocarrier excitation resulting in a decrease in the field inside the ZnO (Fig. 4c, black curve), and after a certain pump fluence, the THG of the film is enhanced with diminishing returns (Fig. 4b).

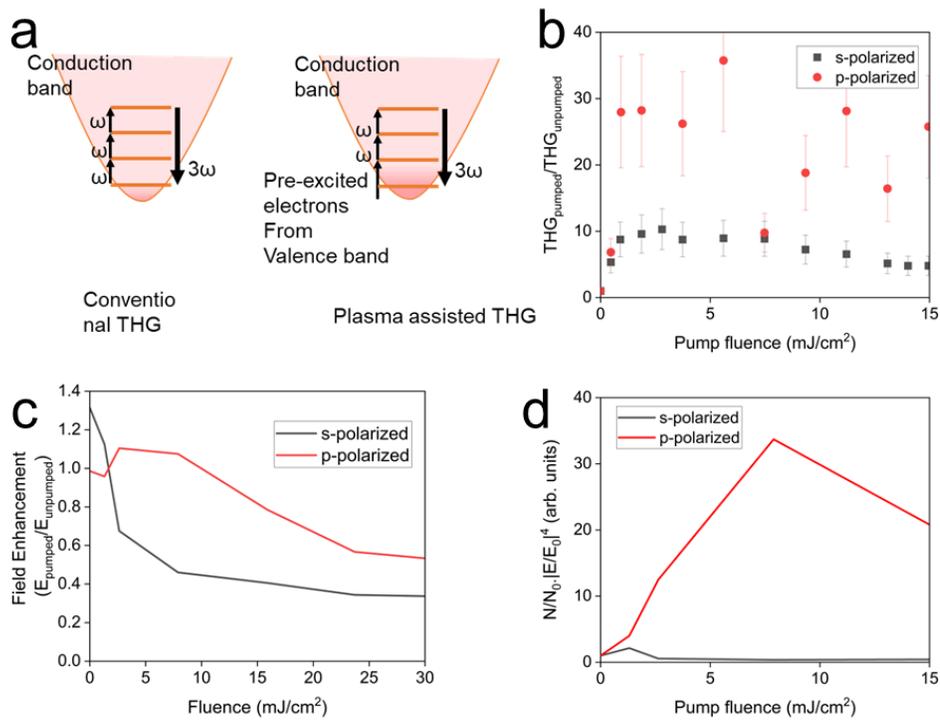

Figure 4 (a) Third harmonic generation mechanism. (b) THG vs the pump fluence for a constant probe (c) Field-enhancement (FIE) versus the pump fluence calculated from the optical properties of zinc oxide (d) Product of the number of free-carriers and the FIE for s- and p-polarized light shows similar trends to experimental enhancement.

In the case of THG from p-polarized light, the increase in photocarrier generation results in a similar increase. At lower pump-powers, the field inside the ZnO also increases as the permittivity of the ZnO decreases, enhancing the field (Fig. 4c). This enhancement of the films with respect to the unpumped state results in a much larger increase in the third harmonic efficiency. In our experiments, we observed an increase in efficiency of up to 50-times for p-polarized light. But after a critical point, which is around 8 mJ/cm$^2$, for this configuration, the increased absorptive losses surpass the field enhancement due to the field discontinuity. The overall enhancement of the films is a function of the number of photoexcited carriers in the conduction band, N, and the 4$^{th}$ power of the pumped versus unpumped field distribution. Employing the number of free carriers computed from the pump-fluence, and the field distribution computed from the permittivities of the various plots, their product (Fig. 4d) shows a good qualitative match with the experimental pump-dependence of the THG enhancement (Fig. 4b). The higher measured enhancement at low pump-fluences observed in the current film is likely a result of it having a different pump absorption, and different dynamic optical properties from the zinc oxide used in the prior study[24].

With the same configuration, enhancements were observed at various other probe wavelengths, showing that this method of enhancement is broadband (See Supporting Information S4).

In conclusion, we demonstrate that harnessing photoexcited carriers can unlock new modes of third harmonic generation, with novel power dependencies, and better efficiency. The method presented here offers enhancements of up to 50 times in the current configuration, but likely can be further optimized for better efficiency. The method is broadband and switchable, and demonstrates a picosecond switching speed, with the prospect of both enhancing and suppressing the nonlinearities by varying the pump power and the pump-probe delay. Whereas prior demonstrations of optically switched nonlinearities have reported a decrease in the efficiency of the higher harmonic signal[50], we are able to achieve a significant increase. Permittivity modulation also offers the opportunity of yet greater nonlinear efficiencies even without the need for patterning or doping.

## Data availability

The source for this study will be available from the authors upon request.

## Acknowledgements

Work performed at the Center for Nanoscale Materials, a U.S. Department of Energy Office of Science User Facility, was supported by the U.S. DOE, Office of Basic Energy Sciences, under Contract No. DE-AC02-06CH11357. The authors acknowledge support by the U.S. Department of Energy, Office of Basic Energy Sciences, Division of Materials Sciences and Engineering under Award DE-SC0017717 (TCO materials growth and characterization), the Office of Naval Research under Award N00014-20-1-2199 (TCO optical characterization), and the Air Force Office of Scientific Research under

## Author Contributions

S.S. conceived the basic idea for this work. S.S., S.J.G., and S.C. performed the COMSOL simulations and the TMM calculations. S.S. fabricated and characterized the device. B.D. and R.D.S. designed the preliminary experimental set-up for the pump-probe spectroscopy. S.S., B.D. and R.D.S. carried out the measurements. S.S., and R.D.S. analyzed the experimental results. A.V.K., V.M.S., A.B. and R.D.S. supervised the research and the development of the manuscript. S.S. wrote the first draft of the manuscript. All co-authors subsequently took part in the revision process and approved the final copy of the manuscript.

## Competing Interests Statement

The authors declare no competing interest.

# Supplementary information

# Third Harmonic Enhancement Harnessing Photoexcitation Unveils New Nonlinearities in Zinc Oxide

Soham Saha[1*], Sudip Gurung[2], Benjamin T. Diroll[1], Suman Chakraborty[1], **Vladimir M. Shalaev[2, 3]**, Alexander V. Kildishev[2, 3], **Alexandra Boltasseva[2, 3] and Richard D. Schaller[1,4]**

[1] *Argonne National Laboratory, Lemont, IL 60439, USA*

[2] *School of Electrical and Computer Engineering, Birck Nanotechnology Center, Purdue University, West Lafayette, IN, USA*

[3] *Purdue Quantum Science and Engineering Institute, Purdue University, West Lafayette, IN, USA*

[4] *Department of Chemistry, Northwestern University, Evanson, IL, USA*

*\*sohamsaha@anl.gov*

*SI 1. Film growth and steady-state characterization*

200 nm of titanium nitride was deposited on magnesium oxide substrates using DC magnetron sputtering at 800°C. A 99.995% pure titanium target of 2 in. the diameter was used. The DC power was set at 200 W. To maintain a high purity of the grown films, the chamber was pumped down to $3\times10^{-8}$ Torr before deposition and backfilled to 5 mTorr during the sputtering process with argon. The throw length was kept at 20 cm, ensuring a uniform thickness of the grown TiN layer throughout the 1 cm by 1cm MgO substrate. After heating, the pressure increased to $1.2\times10^{-7}$ Torr. An argon-nitrogen mixture at a rate of 4 sccm/6 sccm was flowed into the chamber. The deposition rate was 2.2 Å min$^{-1}$.

250 nm of ZnO was deposited on the TiN films by pulsed laser deposition (PLD) with a KrF excimer laser at a pump fluence of 1.5 J/cm$^2$. The chamber was pumped down to $6\times10^{-6}$ T and then backfilled to 3 mTorr with oxygen. The substrate temperature was 45°C, and the growth rate was 6 nm/min.

*SI 2. Optical models of ZnO and TiN*

Spectroscopic ellipsometry and fits: The TiN and the ZnO films were characterized using a J. A. Woollam VASE (Variable Angle Spectroscopic Ellipsometer) at angles of 50 and 70°.

From the visible to the NIR wavelengths, both films can be fitted using a Drude-Lorentz model, where the free carriers are modeled with a Drude oscillator and the net effect of bound electrons with a Lorentz term.

The relative permittivity represented as a complex number, $\varepsilon = \varepsilon_1 + i\varepsilon_2$, is given by the following equation,

$$\varepsilon = \varepsilon_1 + i\varepsilon_2 = \varepsilon_\infty - \frac{A_0}{(\hbar\omega)^2 + iB_0\hbar\omega} + \frac{A_1}{E_1^2 - (\hbar\omega)^2 - iB_1\hbar\omega}$$

where ε1 and ε2 are the real and imaginary parts of the dielectric function, $\hbar$ is the reduced Plank constant, $\hbar\omega$ is the probe energy, and $\varepsilon_\infty$ is an additional offset; $A_0$ is the square of the plasma frequency in electron volts, $B_0$ is the damping factor of the Drude oscillator; $A_1$, $B_1$, and $E_1$ are the amplitude, broadening, and the center energy of the Lorentz oscillator respectively.

*Table S1. Drude-Lorentz Model Parameters for TiN and ZnO*

| Drude Lorentz model parameters | TiN | ZnO |
| --- | --- | --- |
| $\varepsilon_\infty$ | 3.44 | 3.33 |
| $A_0$ (eV)$^2$ | 57.6 | 0.163 |
| $B_0$ (eV) | 0.197 | 0.089 |
| $A_1$ (eV)$^2$ | 119 | 3.48 |
| $B_1$ (eV) | 3.096 | 0.28 |
| $E_1$ (eV) | 5.26 | 3.88 |

*SI 3. Electric field distribution in the zinc oxide film in the unpumped and the pumped states*

Figure S1 shows the field distribution inside the ZnO film at different pump-fluences. For s-polarized light, the component of the electric field is parallel to the interface, and is continuous. Inside the film, the field exponentially decays due to the absorptive losses of the zinc oxide. The field is maximum in the unpumped state, and decreases with increasing pump fluence.

For p-polarized light, with decreasing permittivity, the perpendicular component of the electric field increases inside the ZnO. But after an optimal point, the absorptive losses overcome this effect, and the field amplitude decreases with increasing power.

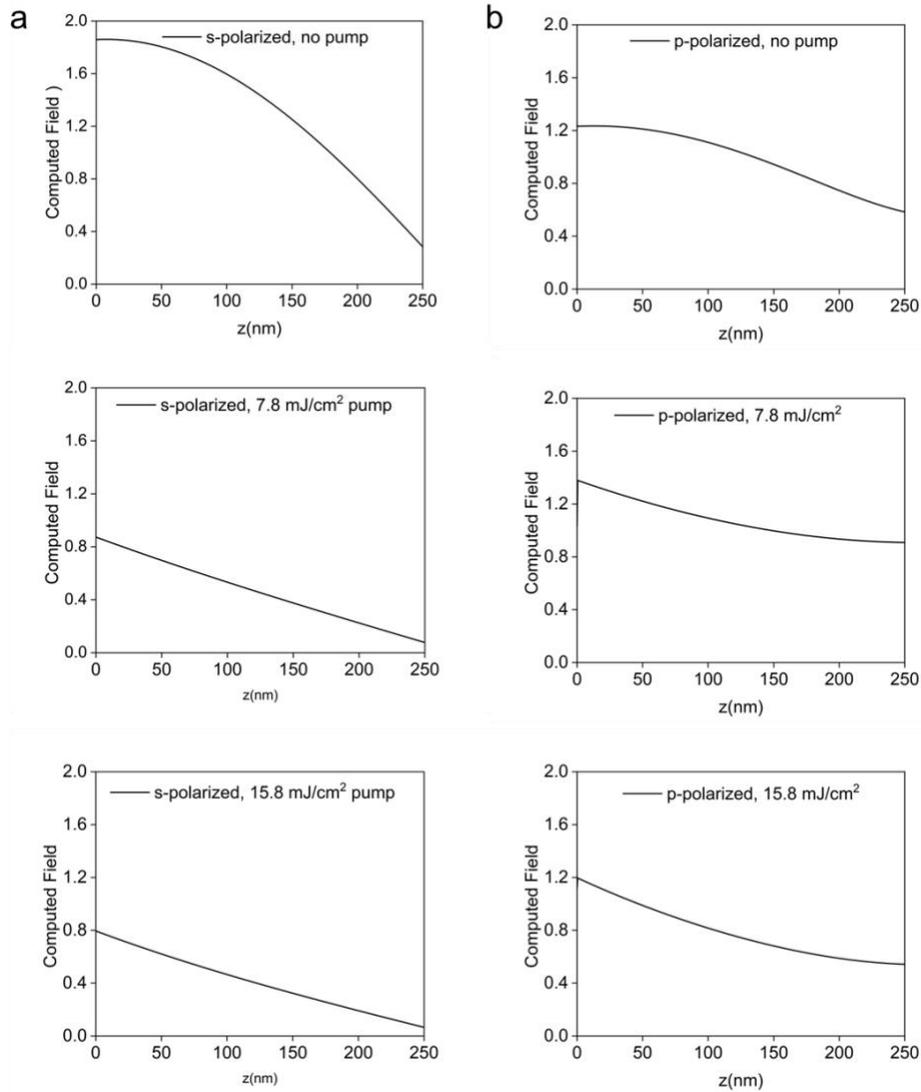

Figure S1 a. Field distribution (parallel to surface) inside the 250 nm ZnO film. z=0 nm is the air-ZnO boundary, and z=250 nm the ZnO-TiN boundary. The field results from the interference between the incident beam with the beam reflected from the TiN-ZnO boundary. The field is highest in the unpumped state, and decreases with increasing pump fluence, as the absorptive losses increase. (b) In p-polarized light, the maximum field occurs at a pump-fluence of 7.8mJ/cm$^2$, where the discontinuity in the perpendicular component of the permittivity is maximum.

After that, a further increase in the fluence results in a decrease in the overall electric field because of absorptive losses, as the film becomes more metallic.

*SI 4. Broadband enhancement of THG enabled by pump*

The utilization of photoexcited carriers enables the enhancement of THG across a broad range of wavelengths, as shown in Fig. S2. An enhancement of around 15-50× is observed in the same experimental configurations, employing the pump.

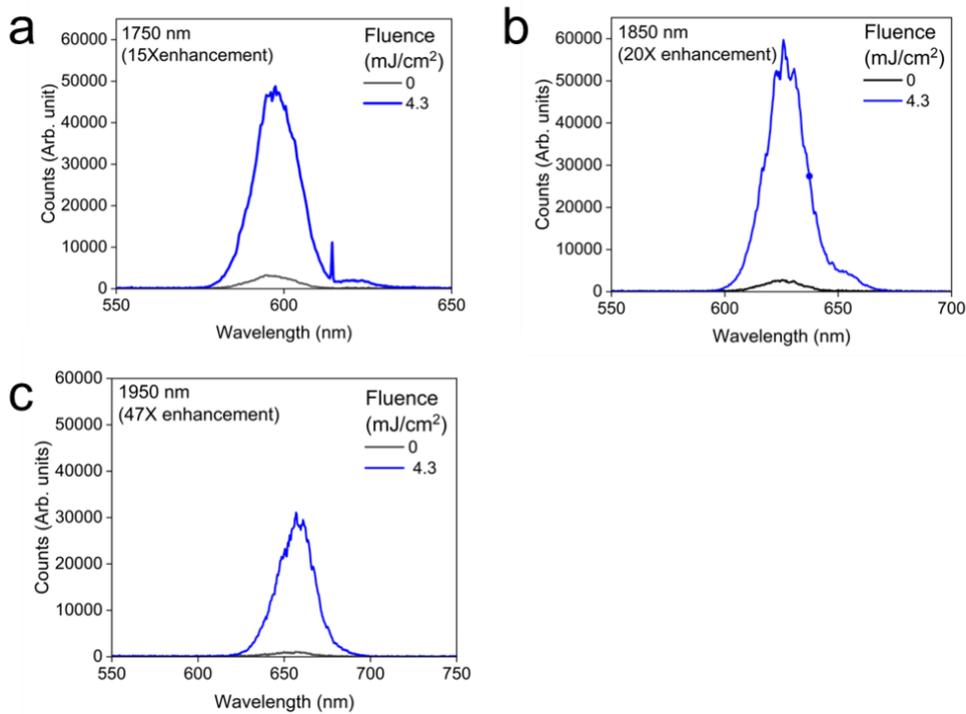

Figure S2 The THG in the unpumped and pumped states of ZnO at a probe wavelength of (a) 1750 nm showing 15× enhancement, (b) 1850 nm showing 20× enhancement, and (c) 1950 nm showing 47× enhancement.